\documentclass[preprint,10pt]{aastex}
\usepackage[utf8]{inputenc}

\def\ha{H$\alpha$}

\def\msun{M$_\odot$}

\def\teff{$T_{\rm eff}$}
\def\logg{$\log g$}
\def\lbol{$L_{\rm bol}$}
\def\lha{$L_{{\rm H}\alpha}$}

\usepackage[normalem]{ulem}
\usepackage[usenames,dvipsnames]{color}

\begin{document}

\title{The Impact of Chromospheric Activity on Observed
Initial Mass Functions}
\author{Keivan G.\ Stassun\altaffilmark{1,2},
Aleks Scholz\altaffilmark{3}, 
Trent J.\ Dupuy\altaffilmark{4}, 
Kaitlin M.\ Kratter\altaffilmark{5}}

\altaffiltext{1}{Department of Physics \& Astronomy, Vanderbilt University, Nashville, TN 37235, USA;
keivan.stassun@vanderbilt.edu}
\altaffiltext{2}{Department of Physics, Fisk University, Nashville, TN 37208, USA}
\altaffiltext{3}{School of Physics \& Astronomy, University of St Andrews, North Haugh, St Andrews, KY16 9SS, United Kingdom}
\altaffiltext{4}{The University of Texas at Austin, Department of Astronomy, 2515 Speedway C1400, Austin, TX 78712, USA}
\altaffiltext{5}{University of Arizona, Steward Observatory, 933 N Cherry Ave, Tucson, AZ, 85721}



\begin{abstract}
Using recently established empirical calibrations for the impact of chromospheric activity
on the radii, effective temperatures, and estimated masses of active low-mass stars and brown dwarfs, 
we reassess the shape of the initial mass function (IMF) across the stellar/substellar boundary
in the Upper Sco star-forming region (age $\sim$5--10 Myr). 
We adjust the observed effective temperatures to warmer values using the observed strength
of the chromospheric H$\alpha$ emission, and redetermine the estimated masses of objects
using pre--main-sequence evolutionary tracks in the H-R diagram. The effect of the 
activity-adjusted temperatures is to shift the objects to higher masses by 3--100\%.
While the slope of the resulting IMF at substellar masses is not strongly changed, the
peak of the IMF does shift from $\approx$0.06 to $\approx$0.11 \msun. Moreover, for objects
with masses $\lesssim$0.2 \msun, the ratio of brown dwarfs to stars changes from $\sim$80\%
to $\sim$33\%. These results suggest that activity corrections are essential for studies
of the substellar mass function, if the masses are estimated from spectral types or
from effective temperatures.
\end{abstract}

\section{Introduction\label{sec:intro}}

A fundamental question in the study of star formation is the
mass spectrum of stars produced in young star clusters---the
initial mass function (IMF). The IMF encodes the
physics from the star formation process, sets the initial conditions
for stellar population modeling, 
and thus informs models of galactic evolution.
Empirical IMFs serve as crucial touchstones
for many aspects of stellar and galactic astrophysics. 


The origin of the IMF and its dependence on environmental factors remains an active area of research \citep{Offner:ppvi}. Substellar objects specifically may provide crucial insight into the IMF as a whole. Consensus is emerging that objects below the Deuterium burning limit ($\sim13M_{\rm Jup}$, \citealt{Baraffe:1998}) can form by the same process as their stellar counterparts \citep{Chabrier:ppvi}. 
In this unified picture of star formation, the formation of low mass objects requires non-linear density fluctuations generated by turbulence. Two other prominent theories for brown dwarf formation are ejection from their gas reservoir due to three-body interactions, and disk fragmentation around a more massive star due to gravitational instability \citep{ReipurthClarke,Bate:2000}. Current observational evidence---relatively massive brown dwarf disks \citep{Ricci:2014}, the continuity of the IMF across the substellar boundary, the discovery of a young brown dwarf binary and a pre-brown dwarf in isolation (\citep{luhman_2009,andre_2012}, as well as the orbital properties of close binaries \citep{Dupuy:2011}---seems to favor a unified turbulent fragmentation theory. 

In the turbulent fragmentation paradigm, turbulence and self-gravity generate a range of density perturbations, which in conjunction with sub fragmentation, produce the full range of stellar masses. Because the brown dwarfs sample the tail of the distribution, it is possible that their abundance is more sensitive to changes in environment than the IMF as a whole. Indeed there is some evidence for different populations of low mass objects from cluster to cluster \citep{scholz_13}, whereas the evidence for a varying IMF in the Milky Way is absent \citep[see, e.g.,][for a review]{bastian2010}. 
As we show in this paper, some of this variation might be due to chromospheric activity contaminating mass determination at the sub-stellar boundary. In contrast, massive elliptical galaxies do show evidence for a bottom heavy IMF \citep{vandokkum_conroy_11}, however it is unlikely that a change in the brown-dwarf-to-star ratio would be detectable given the uncertainties in the overall mass-to-light ratio.

A first step towards understanding the origins of possible IMF variations is a robust calculation of young cluster IMFs. Determining the masses of objects near the substellar boundary in these regions is crucial because it provides a direct probe of the star formation process. More importantly, young clusters allow us to probe further down the mass function because substellar objects are still relatively bright at young ages. 


Empirically determining the IMFs of young clusters is usually done in one of two ways: one can estimate the luminosity function and then convert to a mass function using a theoretical M--L relationship \citep[e.g.,][]{muench_2003}, or one can estimate the masses of the individual stars in the cluster by comparing their position in the H-R diagram with a theoretical isochrone, and then build up the aggregate IMF from those individual masses \citep[e.g.][]{luhman_2003}. In the latter case, the masses can be estimated from effective temperatures (\teff) or 
bolometric luminosities 
(\lbol) or a combination of the two. The distribution of stellar masses is usually found to peak between 0.1--0.5 \msun.
In recent years, the ratio of stars to brown dwarfs 
in individual clusters has frequently been used 
as a simple, quantitative metric to parameterize the 
IMF in the low-mass regime \citep[e.g.,][]{andersen_2008,luhman_2003,scholz_2012}, specifically suited to test the various proposed formation scenarios for brown dwarfs \citep[see][]{scholz_13}.

Importantly, stellar masses inferred from \teff\ measurements
can be significantly underestimated if the effects of magnetic
activity are not taken into account. There is emerging consensus
in the literature that magnetic activity inflates the radii and
suppresses the temperatures of low-mass stars 
\citep[e.g.,][]{Lopez2007,Morales2008,MacDonald2009,Morales2010,Stassun2012}.
An active
star will present a lower \teff\ than expected for its mass;
the stellar mass inferred from that lower \teff\ will in turn 
be lower than the true mass.
In \citet{Stassun2012}, we used mass, radius, and \teff\
measurements for a benchmark sample of active field dwarfs and
eclipsing binaries to derive empirical relationships
between the strength of magnetic activity (as measured by the
strength of the \ha\ chromospheric emission) and the degree
of radius inflation and \teff\ suppression.
We found, for example, that a low-mass star near the H-burning
limit at an age of a few Myr with an \ha-to-bolometric luminosity
ratio of $\log$ \lha/\lbol\ $= -3.5$ (i.e., near the 
chromospheric ``saturation" value of $-3.3$) will have its 
\teff\ decreased by $\approx$7\% (or $\sim$200~K), and thus its
\teff-inferred mass will be a factor of $\sim$2 lower than 
the true mass \citep[cf.\ Fig.\ 7 in][]{Stassun2012}.

In this paper, we perform an initial assessment of the impact of such chromospheric effects on the inferred 
shape of the bottom of the IMF of a young cluster when the stellar masses are determined from observed \teff\ measurements. We use Upper Sco as our first test region for this experiment. 
Being young \citep[5--10 Myr, see][]{Slesnick2008,Pecaut2012},
the region is amenable to a 
study of activity's effects on the inferred masses of low-mass stars
and brown dwarfs at an age where they are still likely to be
magnetically active while no longer being strongly contaminated
by signatures of disk accretion.
Moreover \teff\ and H$\alpha$ equivalent width measurements
are available for the low-mass objects in this region
\citep{Slesnick2008}, allowing their masses and chromospheric activity levels 
to be determined. 

To be clear, it is not our aim in this paper to determine an ``absolute" IMF
for Upper Sco. The observational determination of IMFs is notoriously complicated by
potential observational biases and systematic effects due to, e.g., sample completeness
and contamination, differential reddening, mass segregation, unresolved binarity, etc.
These issues are beyond the scope of this paper.
Rather, our aim is to explore the impact of activity on an observed IMF
independent of these other observational issues, thus providing an assessment of the
differential effect of activity specifically.


In Sec.~\ref{sec:data} we summarize the
data from the literature that we use as well as the relations
that we employ to convert the observed chromospheric activity
measures into estimated stellar mass corrections. 
Sec.~\ref{sec:results} presents our main result that the ratio
of apparent brown dwarfs to stars in a young cluster is 
significantly altered by
these corrections. We conclude with a discussion of some
implications and limitations of this work in Sec.~\ref{sec:disc}.

\section{Data and Methods\label{sec:data}}


\subsection{Study sample}

Our 
sample of young, low-mass stars is taken from the
study of the young ($\sim$5--10 Myr) Upper Sco star-forming region
of \citet{Slesnick2008}. The sample is based on a large-scale
photometric multi-band survey in optical and near-infrared bands. 
\citet{Slesnick2008} present spectroscopy for 243 candidate members
and confirm 145 as bona fide members of the star forming association. 
Combining spectra and photometry, the authors determine luminosities,
\teff, as well as \ha\ equivalent widths, and
also provide an estimate of the object masses and ages based on the
position of objects in the HR diagram. Based on their analysis, the
age of the Upper Sco members is consistent with formation in a single
burst 5\,Myr ago. The mass function 
presented in their paper does not turn over until below 0.05~\msun,
suggesting a possible
overabundance of very low mass stars and brown dwarfs in Upper Sco
compared with other regions.

The \citet{Slesnick2008} sample is ideal for our purposes
because (a) the low-mass population of the region was
characterized with reportedly good completeness 
from $\sim$0.2 \msun\ to well below the nominal
H-burning limit of $\sim$0.08 \msun\
(i.e., the IMF is well sampled near the expected peak
of the stellar mass distribution and below,
(b) the stars are young
enough to be highly magnetically active but old enough that
the fraction of stars with massive disks and high accretion
rates is relatively low, and (c) the \citet{Slesnick2008}
analysis included \teff-inferred stellar masses and reported 
measurements of the \ha\ emission for all of the sources.

To measure the strength of chromospheric activity, we take the \ha\ equivalent widths (EWs) reported by \citet{Slesnick2008} and convert these to \lha/\lbol\ in the same manner as described in \citet{Stassun2012}. In short, we scale the EWs to \ha\ luminosities by multiplying with continuum fluxes at the wavelength of H$\alpha$ taken from the AMES-Dusty models \citep{Allard2000} for low-gravity, solar metallicity objects (\logg\ $= 4.0$) with
\teff\ from 2400 to 4000\,K. The models provide fluxes for a unit area of stellar surface; therefore we also multiply with $4\pi R^2$, where $R$ was determined from $L$ and \teff. This method gives \lha\ for each object, without any assumptions about distance or age. 



\begin{figure}[ht!]
\includegraphics[angle=90,width=\linewidth]{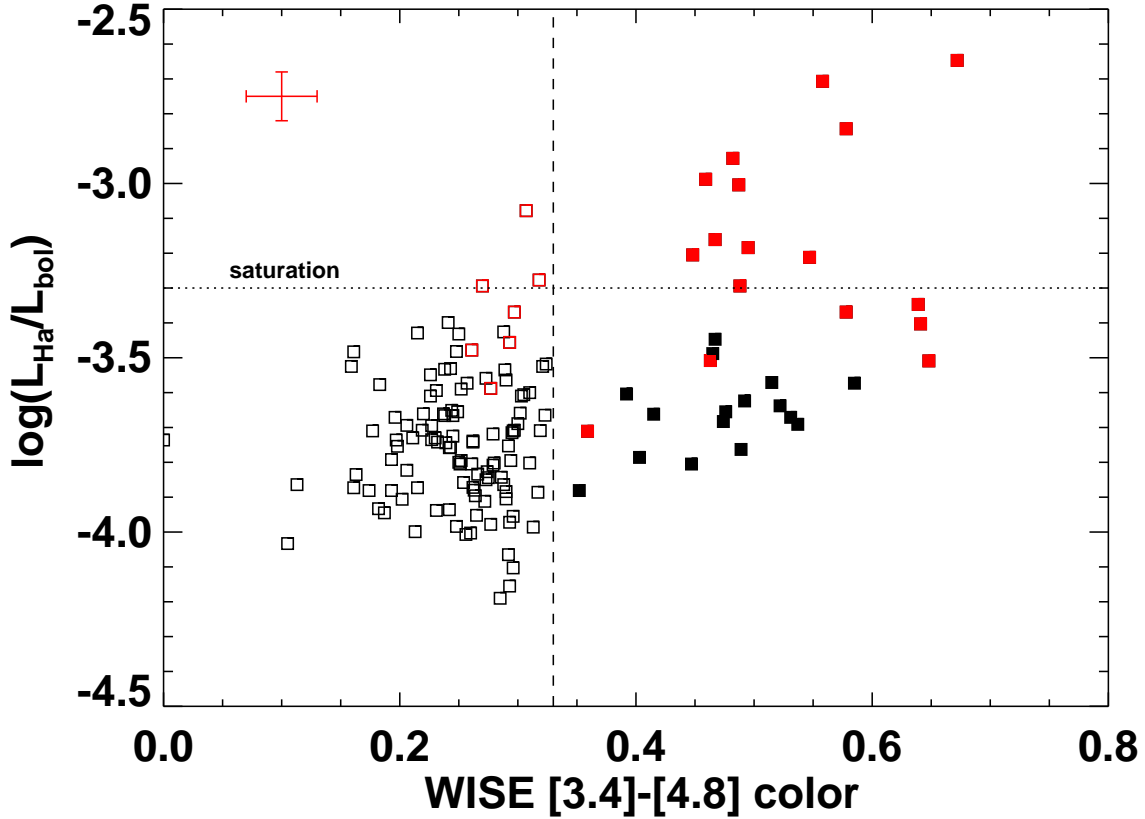}
\caption{Sample of \citet{Slesnick2008} in the
\lha/\lbol\ vs.\ WISE [3.4]$-$[4.8] plane. Stars
with \ha\ emission EW indicative of
active accretion (red symbols) are removed from our study. 
Error bars at upper left represent typical measurement uncertainties.
The vertical line indicates
the [3.4]$-$[4.8] color above which stars 
possess significant excess IR emission indicative of disks
(solid symbols);
these objects are also eliminated from our study.
The horizontal
line indicates the maximum (``saturated") level of \lha/\lbol\ 
expected from chromospheric activity. 
Most of the stars with excessive \ha\ emission (red) are in the upper 
right quadrant,
as expected for stars actively accreting from disks, whereas the 
majority of the sample is in the lower left quadrant as expected for
naked non-accreting stars. 
\label{fig:sample}}
\end{figure}

The Upper Sco study sample of \citet{Slesnick2008} is shown in
Fig.~\ref{fig:sample} in the \lha/\lbol\ vs.\ WISE 
[3.4]$-$[4.8] plane.
A trend of increasing \lha/\lbol\ with increasing WISE color is
apparent, and is due to the presence of strong accretion-induced
\ha\ emission in objects with massive disks as inferred from 
excess infrared (IR) emission in the WISE colors. Since our empirical
relations for correcting the \teff\ due to activity assume that the \ha\
emission is caused by chromospheric activity only, we must eliminate
stars in the sample whose \ha\ emission is potentially contaminated 
by accretion. To that end, we remove stars that satisfy one or both
of the following criteria: (1) \ha\ emission stronger than 20\AA\ EW, 
corresponding to the maximum emission observed in young mid-late M 
stars \citep[cf.\ Fig.\ 4 in][and references therein]{Slesnick2008}, 
and (2) WISE [3.4]$-$[4.8] color larger than 0.33, corresponding to
significant excess IR emission over that expected from bare 
photospheric colors in young mid-late M stars
\citep[see, e.g.,][]{Dawson2013}.

Application of these cuts removes 40 stars (in Fig.~\ref{fig:sample}
these are the red symbols and all objects to the right of the vertical
line), leaving 105 stars in our study sample. Note that none of these
105 sample stars possess \lha/\lbol\ above the nominal chromospheric 
``saturation" limit of $\log$ \lha/\lbol\ $\approx -3.3$ (horizontal
line in Fig.~\ref{fig:sample}), suggesting 
that the retained sample is indeed clean of active accretors.

Note that the removed set (red points and filled points in 
Fig.~\ref{fig:sample}) is larger than (but inclusive of) the set
identified as accretors by \citet{Slesnick2008} because here we have 
conservatively also flagged objects that show clear signatures of 
massive disks from the {\it WISE} infrared data that were not available
at the time of the \citet{Slesnick2008} study. This allows for the possibility
of objects that may not have been observed to be actively accreting at the
epoch when the \ha\ spectra were obtained but that may be affected by 
accretion at other times nonetheless. Note also that our removal 
of the reddest objects in the {\it WISE} passbands does not imply removal
of the lowest mass objects, as the {\it WISE} colors are primarily probing the 
presence of circumstellar disks, not the photospheric \teff. We have checked 
that the removed set does not represent a distinct region of the IMF; a 
two-sided K-S test of the masses of the removed set versus the masses of
the retained set gives a probability of 32\% that they are drawn from the
same parent sample.

\subsection{Stellar masses and empirical activity corrections}

To ascertain the effect of the observed activity on the inferred masses
of the sample stars, we use the empirical relation between \lha/\lbol\ 
and \teff\ suppression determined by \citet{Stassun2012}. Specifically,
we adjust the observed \teff\ upward according to Eq.~1 in
\citet{Stassun2012}:
\begin{equation} \label{eq}
\Delta T_{\rm eff} / T_{\rm eff} = m_T \times 
\left( \log L_{H\alpha}/L_{\rm bol} + 4 \right) + b_T 
\end{equation}
where the relation fit coefficients from Table~1 in \citet{Stassun2012}
are $m_T = -4.71 \pm 2.33$ and $b_T = -4.4 \pm 0.6$, in percent units.

For consistency with the methods employed by \citet{Slesnick2008},
we adopt their reported luminosities as is. We also adopt
their reported spectroscopic \teff\ determinations 
and then adjust them using Eq.~\ref{eq} above. 

We infer the adjusted masses of the stars by
interpolating in \teff\ and $L$ using the same pre--main-sequence
stellar evolutionary models of \citet{DM97} as adopted by
\citet{Slesnick2008}. 
Note that the empirical corrections of \citet{Stassun2012} do not
depend on any particular choice of pre--main-sequence models, because the
corrections depend only on the observed \teff\ and \ha\ emission. However,
the masses inferred from the \teff\ and $L$ certainly do depend on 
the choice of models, hence our use here of the same models originally adopted 
by \citet{Slesnick2008} in order to compare the adjusted masses to the 
originally reported ones.
Finally, to ensure an accurate comparison
between the adjusted masses determined here and the masses
originally inferred by \citet{Slesnick2008}, we checked that
we were able to reproduce the original masses from 
\citet{Slesnick2008} using their originally reported \teff\ 
and $L$. In all cases we reproduced their originally reported
masses to within 1\%, implying that we are effectively adopting
the same interpolation method on the DM97 evolutionary tracks.

\section{Results\label{sec:results}}

\begin{figure}[ht!]
\begin{center}
\includegraphics[width=0.5\linewidth]{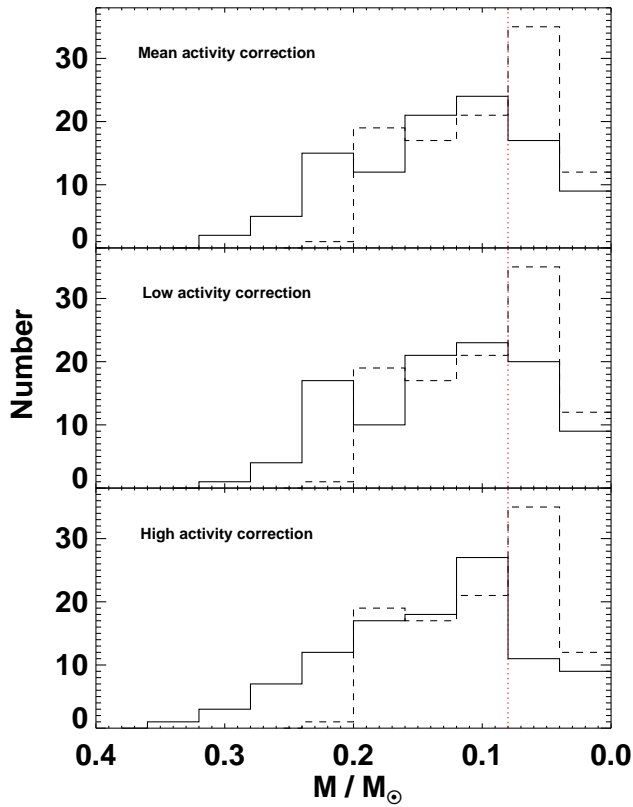}
\end{center}
\caption{Initial mass function of Upper Sco stars from \citet{Slesnick2008}
before (dashed histograms) and after application of activity corrections
(solid histograms) according to Eq.~\ref{eq}. The top panel uses the mean
coefficients in Eq.~\ref{eq}, whereas the middle and bottom panels use the
1$\sigma$ low and high coefficients, respectively.
The vertical line in each panel denotes the nominal brown-dwarf mass limit 
of 0.08 \msun.
The ratio of BDs to stars is lower in the activity-corrected 
IMFs at very high statistical significance (see text).
\label{fig:imf}}
\end{figure}

The result of adjusting the masses of the study sample stars according
to the observed \ha\ emission is shown in Fig.~\ref{fig:imf}. The 
masses are binned by 0.04 \msun, and in each panel the original masses
from \citet{Slesnick2008} are represented by the solid histogram. 
For comparison, the activity-adjusted masses are represented as dashed 
histograms. Because the activity correction coefficients in Eq.~\ref{eq}
have associated uncertainties, we show in the different panels of
Fig.~\ref{fig:imf} the activity-adjusted mass histograms that result from
adopting the coefficients at their mean values (top panel) or at their 
1$\sigma$ low or high values (middle and bottom panels).
Visually, the original IMF from \citet{Slesnick2008} appears to be 
shifted to systematically lower masses, as expected since the activity
corrections act systematically to adjust the \teff\ to higher values
corresponding to higher masses. Considering the sample on an
object-by-object basis, the smallest change in mass in 3\%
while the largest change in mass is 102\%.

If we quantify the shift in the overall sample by comparing
the modal values of the distributions in Fig.~\ref{fig:imf}, we find the
following modal values for the mean correction, low correction, and high
correction cases, respectively (in units of \msun): 
$0.111 \pm 0.007$,
$0.104 \pm 0.006$, and
$0.119 \pm 0.007$.
In comparison, the modal value of the uncorrected distribution is
$0.056 \pm 0.004$ \msun.
In other words, the activity corrected modal mass is consistently
higher than the uncorrected modal mass, and the difference is in all cases
statistically significant at $> 6\sigma$.
In addition, a Wilcoxon rank-sum test yields a probability 
of $<10^{-5}$ in all cases for the null hypothesis of no
difference in the sample medians before and after application
of the activity correction.

However, the intrinsic breadth of the IMF may make shifts in the 
IMF peak difficult to quantify with the modal value; indeed, the IMF appears to be
broadly ``flat" in the vicinity of the peak 
\citep[][and see Fig.~\ref{fig:imf}]{Slesnick2008}.
Thus a number of previous studies have instead used the ratio of
brown dwarfs to stars (or vice versa) as an alternative metric for 
characterizing the IMF \citep{luhman_2003,andersen_2008,scholz_2012}.
In the case of the IMFs shown in Fig.~\ref{fig:imf}, the BD-to-star
ratio for the mean correction, low correction, and high correction
cases, respectively, are
$0.33 \pm 0.07$,
$0.38 \pm 0.08$, and
$0.24 \pm 0.06$,
where we define BDs as objects with inferred masses $< 0.08$ \msun.
In comparison, the BD-to-star ratio for the uncorrected distribution
is $0.81 \pm 0.16$.
The uncertainties in the above ratios are based on simple 
Poisson errors based on the numbers of objects in the sample
above and below 0.08 \msun.
By this simple statistical treatment, 
the activity-corrected and uncorrected IMFs differ in their
BD-to-star ratios at the level of $\sim$3$\sigma$.
However, using a proper statistical test for differences in two
sample proportions\footnote{We use the chi-square test for
equal proportions as implemented in the {\sc R} statistics
package.} \citep{Newcombe1998}, we find a probability 
of $<10^{-6}$ in all cases for the null hypothesis that the
proportions  are the same before and after application of the
activity correction.

\section{Discussion and Conclusions\label{sec:disc}}


In our earlier work \citep{Stassun2012}, we discussed the fact that we saw no obvious discontinuity in our \teff\ suppression relations at the fully convective mass boundary ($M \sim 0.3$\,\msun).  Indeed, our empirical correction seemed to solve the \teff\ reversal problem for the fully convective, young brown dwarf eclipsing binary 2M0535$-$05.  Our study sample in this current work comprises low mass stars and brown dwarfs that have likely not begun hydrogen fusion yet, e.g., \citet{bur01} shows that the core temperature of a 0.2\,\msun\ star does not reach the critical value of $3\times10^6$\,K until $\approx$11\,Myr.  In that sense, these young stars are more like the brown dwarf eclipsing binary than field stars of similar temperature.  One key feature of \teff\ suppression in field stars is that \lbol\ does not change with activity because the energy output is purely set by fusion reaction rate in the core.  Given that we do not yet fully understand the mechanism behind \teff\ suppression, it is possible that the empirical relation derived in the field would not apply to pre--main-sequence stars of similar \teff.

Our correction may resolve an issue raised by the analysis in \citet{Slesnick2008}. Their results suggest that Upper Sco ``contains relatively higher numbers of very low-mass stars and brown dwarfs compared with other star forming regions." The slope of the low-mass IMF derived by \citet{Slesnick2008} is higher than usual ($\alpha = -1.13$ compared with $0.6$ in most other regions, \citet{scholz_2012}). While the mass distribution begins to turn over around 0.1 \msun\ in other regions, \citet{Slesnick2008} report a secondary peak at 0.05 \msun. All this would indicate an overabundance of substellar objects in Upper Sco, which is not confirmed by other groups \citep{Dawson2011}. \citet{Slesnick2008} speculate that this overabundance may be related to the presence of OB stars in Upper Sco. We have shown above that after applying the correction, the peak of the mass function shifts to 0.1 \msun\ or higher, and the
secondary peak disappears. The star-to-brown-dwarf ratio increases as well, and is now consistent with values for other regions \citep[see][for a review]{scholz_13}, which have been derived without using \teff. Thus, after taking into account the effect of magnetic activity on the estimated masses, the low-mass IMF in Upper Sco no longer appears abnormal.

Our current work has not accounted for possible contamination by unresolved binary stars, however we expect these to have a minimal impact on our correction for the mass-radius anomaly. Generically, unidentified binaries can confuse determination of stellar properties because of mismatches between luminosity, effective temperature and \logg. Close, tidally interacting binaries are also known to persist as chromospherically active and to show enhanced \ha\ emission. However, stars in this mass regime have a lower total binary fraction than higher mass stars, closer to 30\% \citep{Duchene_Kraus2013}. More importantly, the parameter space in which to ``hide" a companion is small. Companions close enough to induce activity are statistically biased towards more equal masses,  making them likely to show up as double lined, spectroscopic binaries \citep{Reid_Gizis1997}. For wider systems, nearly equal mass binaries---the easiest to detect---produce the most severe errors in determining stellar properites: these systems would show higher luminosities at a given \teff. Thus one would infer a larger radius, but not an unusually high \lha/\lbol. These systems would only degrade the correlation we identify here.

We have demonstrated that the mass distribution in the very low mass regime is significantly altered by the effects of magnetic activity. If masses are estimated from \teff\, they can be systematically underestimated, which would influence the derived IMF and especially the star-to-brown-dwarf ratio. When masses are estimated from luminosity, this issue is avoided, however other uncertainties arise. During the pre--main-sequence contraction stars drop sharply in luminosity, and so the estimated masses depend strongly on the assumed age, which is typically uncertain by several Myr. Luminosities are also affected by uncertainties in extinction and distance. 
For comparison, a 10\% error in a cluster distance manifests as a 20\% error in luminosity, which in turn biases masses by roughly 10--20\% at the substellar boundary,
and by up to 50\% below the substellar boundary where deuterium burning effects
become particularly important. 
Similarly, a change of 2 Myr ($\approx$20\%) in assumed age for a region
like Upper Sco can translate to 20--50\% change in mass.
Ideally IMF measurements for young clusters should be estimated by multiple techniques to combat these uncertainties.

To our knowledge, the Upper Sco dataset published by \citet{Slesnick2008} is the only large sample of young low-mass objects with \teff\ and \ha\ measurements for every object. This makes it possible to estimate masses from \teff\ while correcting for the influence of magnetic activity. To extend this type of analysis and test the IMF in diverse regions, it would be desirable to obtain activity measurements (e.g., \ha\ EWs) for large samples of young very low mass objects in other star forming regions.

An ancillary impact of accurate mass determinations in young
clusters is the validation of atmosphere models for comparison with young exoplanets, as
there remains a need to understand how spectral class maps to mass and age in isolated objects in order to make comparison models for massive exoplanets.
Most importantly, 
measuring the tail of the IMF in young clusters is vital for our understanding of star formation, and in particular the role of turbulence in producing cluster-cluster variations. If changes to the IMF are most dramatic at lower masses, accurate measurements in young clusters are most important. 
Our work here suggests that unmodeled physics such as magnetic activity can substantially influence the low mass end of the IMF.

\acknowledgments
KGS acknowledges NSF grants PAARE AST-0849736 and
AAG AST-1109612. The work of AS for this paper has been supported by the Science Foundation Ireland through the grant 10/RFP/AST278.
We thank the anonymous referee for helpful suggestions that improved the paper.


\begin{thebibliography}{}

\bibitem[Allard et al.(2000)]{Allard2000} Allard, F., Hauschildt, 
P.~H., \& Schweitzer, A.\ 2000, \apj, 539, 366

\bibitem[Andersen et al.(2008)]{andersen_2008} Andersen, M., Meyer, 
M.~R., Greissl, J., \& Aversa, A.\ 2008, \apjl, 683, L183 

\bibitem[Andr{\'e} et al.(2012)]{andre_2012} Andr{\'e}, P., 
Ward-Thompson, D., \& Greaves, J.\ 2012, Science, 337, 69 

\bibitem[Baraffe et 
al.(1998)]{Baraffe:1998} Baraffe, I., Chabrier, G., Allard, F., \& Hauschildt, P.~H.\ 1998, \aap, 337, 403

\bibitem[Bastian et 
al.(2010)]{bastian2010} Bastian, N., Covey, K.~R., \& Meyer, M.~R.\ 2010, \araa, 48, 339

\bibitem[Bate(2000)]{Bate:2000} Bate, M.~R.\ 2000, \mnras, 314, 
33

\bibitem[{Burrows {et~al.}(2001)Burrows, {Hubbard}, {Lunine}, \&
  {Liebert}}]{bur01}
Burrows, A., {Hubbard}, W.~B., {Lunine}, J.~I., \& {Liebert}, J. 2001, Reviews
  of Modern Physics, 73, 719
  
\bibitem[Chabrier et al.(2014)]{Chabrier:ppvi} Chabrier, G., 
Johansen, A., Janson, M., \& Rafikov, R.\ 2014, arXiv:1401.7559
  
\bibitem[D'Antona \& Mazzitelli(1997)]{DM97} 
D'Antona, F., \& Mazzitelli, I.\ 1997, \memsai, 68, 807

\bibitem[Dawson et al.(2011)]{Dawson2011} Dawson, P., Scholz, A., 
\& Ray, T.~P.\ 2011, \mnras, 418, 1231 

\bibitem[Dawson et al.(2013)]{Dawson2013} Dawson, P., Scholz, A., 
Ray, T.~P., et al.\ 2013, \mnras, 429, 903 

\bibitem[Duch{\^e}ne \& {Kraus}(2013)]{Duchene_Kraus2013} Duch{\^e}ne, G., \& Kraus, A.\ 2013, \araa, 51, 269

\bibitem[Dupuy 
\& Liu(2011)]{Dupuy:2011} Dupuy, T.~J., \& Liu, M.~C.\ 2011, \apj, 733, 122

\bibitem[L{\'o}pez-Morales(2007)]{Lopez2007} L{\'o}pez-Morales, 
M.\ 2007, \apj, 660, 732 

\bibitem[Luhman et al.(2003)]{luhman_2003} Luhman, K.~L., Stauffer, 
J.~R., Muench, A.~A., et al.\ 2003, \apj, 593, 1093 

\bibitem[Luhman et al.(2009)]{luhman_2009} Luhman, K.~L., Mamajek, 
E.~E., Allen, P.~R., Muench, A.~A., 
\& Finkbeiner, D.~P.\ 2009, \apj, 691, 1265 

\bibitem[MacDonald 
\& Mullan(2009)]{MacDonald2009} MacDonald, J., \& Mullan, D.~J.\ 2009, \apj, 700, 387 

\bibitem[Morales et al.(2010)]{Morales2010} Morales, J.~C., 
Gallardo, J., Ribas, I., et al.\ 2010, \apj, 718, 502 

\bibitem[Morales et 
al.(2008)]{Morales2008} Morales, J.~C., Ribas, I., \& Jordi, C.\ 2008, \aap, 478, 507 

\bibitem[Muench et al.(2003)]{muench_2003} Muench, A.~A., Lada, 
E.~A., Lada, C.~J., et al.\ 2003, \aj, 125, 2029 

\bibitem[Newcombe(1998)]{Newcombe1998} Newcombe, R.~G.\ 1998, 
Statistics in Medicine, 17, 873

\bibitem[Offner et al.(2013)]{Offner:ppvi} Offner, S.~S.~R., Clark, 
P.~C., Hennebelle, P., et al.\ 2013, arXiv:1312.5326

\bibitem[Pecaut et al.(2012)]{Pecaut2012} Pecaut, M.~J., Mamajek, 
E.~E., \& Bubar, E.~J.\ 2012, \apj, 746, 154 

\bibitem[Reid \& Gizis(1997)]{Reid_Gizis1997} Reid, I.~N., \& Gizis, J.~E.\
1997, \aj, 114, 1992

\bibitem[Reipurth 
\& Clarke(2001)]{ReipurthClarke} Reipurth, B., \& Clarke, C.\ 2001, \aj, 122, 432

\bibitem[Ricci et al.(2014)]{Ricci:2014} Ricci, L., Testi, L., 
Natta, A., et al.\ 2014, arXiv:1406.0635

\bibitem[Scholz et al.(2012)]{scholz_2012} Scholz, A., Muzic, K., 
Geers, V., et al.\ 2012, \apj, 744, 6 

\bibitem[Scholz et al.(2013)]{scholz_13} Scholz, A., Geers, V., 
Clark, P., Jayawardhana, R., \& Muzic, K.\ 2013, \apj, 775, 138 

\bibitem[Slesnick et al.(2008)]{Slesnick2008} Slesnick, C.~L., 
Hillenbrand, L.~A., \& Carpenter, J.~M.\ 2008, \apj, 688, 377 

\bibitem[Stassun et al.(2012)]{Stassun2012} Stassun, K.~G., 
Kratter, K.~M., Scholz, A., \& Dupuy, T.~J.\ 2012, \apj, 756, 47

\bibitem[van Dokkum 
\& Conroy(2011)]{vandokkum_conroy_11} van Dokkum, P.~G., \& Conroy, C.\ 2011, \apjl, 735, L13A

\end{thebibliography}
\end{document}